\documentclass[conference, 10pt]{IEEEtran}
\IEEEoverridecommandlockouts
\usepackage[table]{xcolor}
\usepackage{latexsym}
\usepackage{graphicx}
\usepackage{amsfonts,amssymb,amsmath}
\usepackage{cite}
\usepackage{textcomp}
\usepackage{gensymb}
\usepackage{siunitx}
\usepackage{xcolor}
\usepackage{tikz}
\usepackage{pgfplots}
\pgfplotsset{compat=newest}
\pgfplotsset{plot coordinates/math parser=false}
\usetikzlibrary{plotmarks,shapes,patterns,decorations.pathreplacing,backgrounds,calc,arrows,arrows.meta,spy,matrix,shadows,trees,positioning}
\usepgfplotslibrary{patchplots,groupplots}
\usepackage{tikzscale}
\usepackage{tikz-qtree}
\usepackage{etoolbox}
\usepackage{multirow}
\usepackage{algorithm,algorithmic}
\usepackage{setspace}
\usepackage[font=scriptsize]{subcaption}
\usepackage[font=scriptsize]{caption}
\usepackage{float}
\usepackage{url}
\usepackage{xspace}
\usepackage{siunitx}
\usepackage{hhline}
\usepackage{afterpage}

\def\nb0{{\mathbf{0}}}
\def\nb1{{\mathbf{1}}}

\usepackage[acronyms,nonumberlist,nopostdot,nomain,nogroupskip]{glossaries}

\newacronym{pblv}{pBLV}{people with blindness and low vision}
\newacronym{who}{WHO}{world health organization}
\newacronym{gnss}{GNSS}{global navigation satellite system}
\newacronym{imu}{IMU}{inertial measurement unit}
\newacronym{ai}{AI}{artificial intelligence}
\newacronym{llm}{LLM}{large language model}
\newacronym{eta}{ETA}{electronic travel aid}
\newacronym{oai}{OAI}{open air interface}
\newacronym{ran}{RAN}{radio access network}
\newacronym{sdr}{SDR}{software-defined radio}
\newacronym{usrp}{USRP}{universal software radio peripheral}
\newacronym{cots}{COTS}{commercial off-the-shelf} 
\newacronym{5gc}{5GC}{5G core}
\newacronym{sim}{SIM}{subscriber identity module}
\newacronym{3gpp}{3GPP}{the 3rd generation partnership project}
\newacronym{bwp}{BWP}{bandwidth part}
\newacronym{gnb}{gNB}{next-generation node base}
\newacronym{gpu}{GPU}{graphics processing unit}
\newacronym{qos}{QoS}{quality of service}
\newacronym{ue}{UE}{user equipment}
\newacronym{cdf}{CDF}{cumulative distribution function}
\newacronym{hpc}{HPC}{high performance computing}
\newacronym{vlm}{VLM}{vision-language model}
\newacronym{gps}{GPS}{global positioning system}
\newacronym{rtp}{RTP}{real-time transport protocol}
\newacronym{pmf}{PMF}{probability mass function}
\newtoggle{conf}
\toggletrue{conf}

\def\BibTeX{{\rm B\kern-.05em{\sc i\kern-.025em b}\kern-.08em T\kern-.1667em\lower.7ex\hbox{E}\kern-.125emX}}

\newcommand{\VISION}{VIS$^4$ION\xspace}
\definecolor{vividviolet}{rgb}{0.65, 0.0, 1.0}

\begin{document}

\title{5G Edge Vision: Wearable Assistive Technology for People with Blindness and Low Vision}

\author{
\IEEEauthorblockN{Tommy Azzino$^{*}$, Marco Mezzavilla$^{*}$, Sundeep Rangan$^{*}$, Yao Wang$^{*}$, John-Ross Rizzo$^{*\ \dagger}$}

\IEEEauthorblockA{$^{*}$NYU Tandon School of Engineering, Brooklyn, NY, USA - \{ta1731, mezzavilla, srangan, yw523\}@nyu.edu}
$^{\dagger}$Department of Rehabilitation Medicine, NYU Langone, New York, NY, USA - \{johnrossrizzo\}@gmail.com
\vspace{-3mm}

}

\maketitle
\begin{abstract}

In an increasingly visual world, \gls{pblv} face substantial challenges in navigating their surroundings and interpreting visual information. From our previous work, \VISION is a smart wearable that helps \gls{pblv} in their day-to-day challenges. It enables multiple microservices based on \gls{ai}, such as visual scene processing, navigation, and vision-language inference. These microservices require powerful computational resources and, in some cases, stringent inference times, hence the need to offload computation to edge servers.
This paper introduces a novel video streaming platform that improves the capabilities of \VISION by providing real-time support of the microservices at the network edge. When video is offloaded wirelessly to the edge, the time-varying nature of the wireless network requires adaptation strategies for a seamless video service. We demonstrate the performance of our adaptive real-time video streaming platform through experimentation with an open-source 5G deployment based on \gls{oai}. The experiments demonstrate the ability to provide microservices robustly in time-varying network conditions.

\end{abstract}


\IEEEpeerreviewmaketitle

\glsresetall
\section{Introduction}

There are 39 million blind and 246 million people with low vision worldwide, according to \gls{who} \cite{WHO282}. While therapeutic advances are being developed for a handful of conditions, there are a multitude of etiologies that result in severe visual disability \cite{bourne2013causes}, and the prevalence of many of these conditions is increasing. Impaired vision constrains mobility, 
inevitably leading to problems with unemployment and quality of life \cite{sherrod2014association},
both of which limit psychosocial well-being.

\begin{figure}[t]
    \centering
    \includegraphics[width=0.8\linewidth]{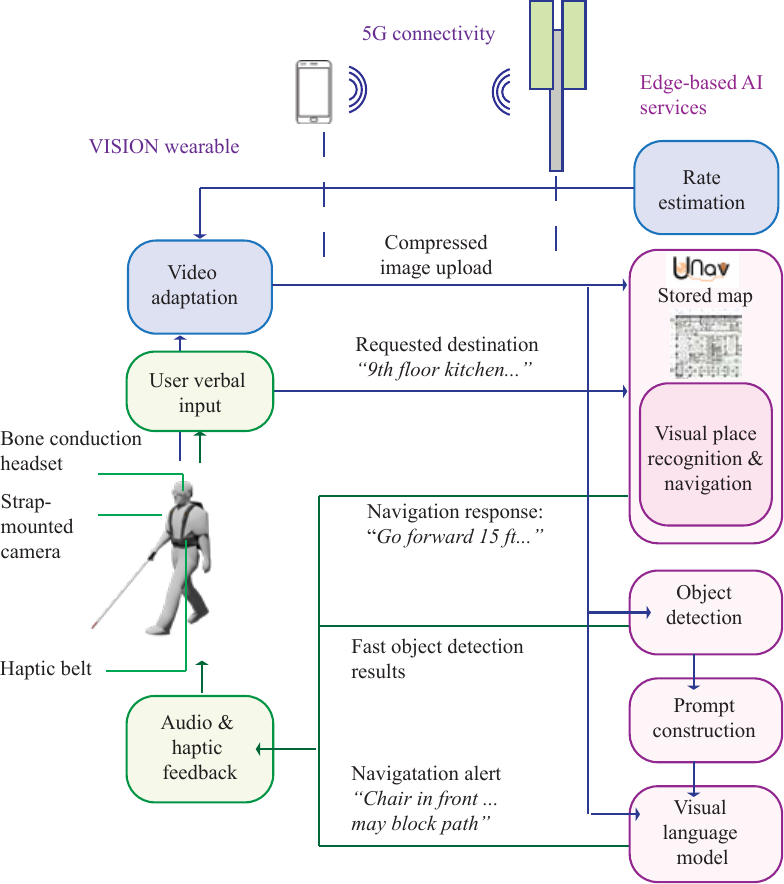}
    \caption{The proposed \VISION platform for people with blindness and low vision provides several powerful edge-based AI microservices. These services all use video captured on a strap mounted camera on the wearable.}
    \label{fig:arch}
    \vspace{-4mm}
\end{figure}

To address the challenges of \gls{pblv}, we have developed \VISION (visually impaired smart service system for spatial intelligence and onboard navigation) \cite{arc-1, arc-8} -- a discreet and ergonomic wearable equipped with miniaturized sensors, including cameras, microphones, \gls{gnss} receivers, and \glspl{imu}. Real-time feedback is provided through a binaural bone conduction headset and an optional reconfigurable waist strap turned haptic interface, as depicted on the left of Fig.~\ref{fig:arch}.

 \VISION offers multiple microservices based on \gls{ai} for \gls{pblv} with visual data as input. These include object detection for obstacle avoidance, \gls{llm}-based audio assistance with visual data, and vision-based localization and navigation. Ideally, these microservices run on the edge, where powerful computational resources enable faster inference and deployment of larger models for optimal performance. However, edge offloading requires seamless wireless connectivity to continuously upload and process video and audio data.

A critical challenge in developing \VISION\ with edge offloading -- and the focus of this work -- is wireless connectivity. Wireless deployment of vision-based edge services suffers from the time-varying nature of network conditions \cite{azzino}. Wireless connectivity will vary with the location of the user relative to the access point or base station, as well as with the network load. Therefore, the bitrate and processing frequency of the video frames will need to be adjusted based on such conditions. Most of the previous work on video adaption for edge-based services has focused on object detection \cite{intro_ta_1}. On the contrary, the \VISION system requires the adaptation of video that caters to multiple services with different \gls{qos} requirements.

The paper presents several important contributions to address these challenges and advance the set of features supported by \VISION:
\begin{itemize}
    \item We develop a full video streaming platform to support real-time fruition of the envisioned microservices for \gls{pblv}. Implementation supports concurrent execution of the \gls{ai}-based microservices over a pool of available \gls{gpu} resources. 
    \item We implement a method to estimate the available link rate using our REBERA algorithm \cite{REBERA}. Based on the link rate estimate, a heuristic algorithm is proposed to adapt the video across different services.
    \item We evaluate the proposed adaptive video streaming platform using an open-source 5G testbed based on \gls{oai} in an indoor lab environment. 5G link conditions are manipulated by changing the available bandwidth to emulate a network under time-varying load.
\end{itemize}


\section{\VISION with 5G Connectivity
and Video Adaptation}
\label{sec:vision}

\subsection*{\VISION microservices architecture}
\label{subsec:microservices}

\VISION wearable is wirelessly connected to a set of powerful edge microservices based on \gls{ai}, including visual scene processing, real-time localization and navigation, and audio assistance through \glspl{vlm}. 
The architecture for these services is schematically depicted in Fig.~\ref{fig:arch}.
The primary input for all services is video, which is captured from the wearable strap-mounted
cameras for image stability. The captured video is compressed and uploaded wirelessly to the edge.  
In this work, we explore 5G connectivity, although any wireless connectivity can be used, including Wi-Fi. As we explain below, the uplink rate can vary, and adaptation of the video compression is critical.

Several edge microservices use the uploaded video data. The first service is \textbf{object detection} using \emph{YoLo}. Building on our previous work in \cite{azzino}, our video streaming platform allows real-time high frame rate object detection using one of the most recent YoLo models, YoLoV7 \cite{yolov7}. In particular, we use \emph{YoLoV7-w6} which supports inference on higher resolution images compared to the basic YoLoV7 model.
In order to support fast responses for immediate and dangerous obstacles, the object detection results
are generally reported at a fast frame rate and then rendered to the user via haptic or audio feedback.
Our analysis in \cite{azzino} suggested that object detection should be as fast as 30 fps to allow a
response time of 100 ms when including video upload, inference, and feedback time.

The second microservice offered is \textbf{UNav} \cite{yang2022unav}. UNav is an \gls{ai}-based software that creates infrastructure-free, camera-based digital twins or 3D maps of complex indoor and outdoor environments, supporting wayfinding to close or far-range destinations through audio and haptic user commands. Note that the commonly used \gls{gps} would not work indoors and has a relatively low accuracy.  
UNav relies on \emph{visual place recognition} \cite{lowry2015visual} where the features of the query
image are compared against a library of pre-stored features to locate the user. The location information
can then be used to provide navigation instructions.  
For example, as shown in Fig.~\ref{fig:arch}, the user could verbally request a destination such as the kitchen, and UNav first finds the user's current location from the current video frame and then provides a text-based navigation suggestion such as ``go forward" a certain distance. The navigation
suggestions are then converted to audio using text-to-speech tools and delivered to the user through the headset.   

Lastly, our platform supports real-time audio assistance using a \textbf{\gls{vlm}} called \emph{InstructBLIP} \cite{instructblip}. InstructBLIP is an \gls{llm}-based \gls{vlm} for comprehensive scene understanding and textual descriptions. It can generate descriptive output text based on the input prompt and image. Specifically, InstructBLIP begins by capturing high-level visual representations of the image. Then, the input prompt and visual features are used to generate contextualized information through a \gls{llm}. Specifically, InstructBLIP works in association with a \gls{llm} known as Vicuna-13B \cite{vicuna2023} to generate the final output text.

Each service has different computational and feedback frequency requirements. As described above, 
object detection inference must support a high frame rate to capture suddenly appearing objects that might pose a risk to \gls{pblv}, and therefore feedback must be received within a strict delay budget. On the other hand, navigation and audio assistance require complex models and can often be executed with lower inference frequency. For example, localization and navigation may be sufficient at 1 feedback per second. The same is true with audio assistive on the general scene composition and relatively far obstacles.

\begin{figure*}[t!]
\centering
\includegraphics[width=0.83\textwidth]{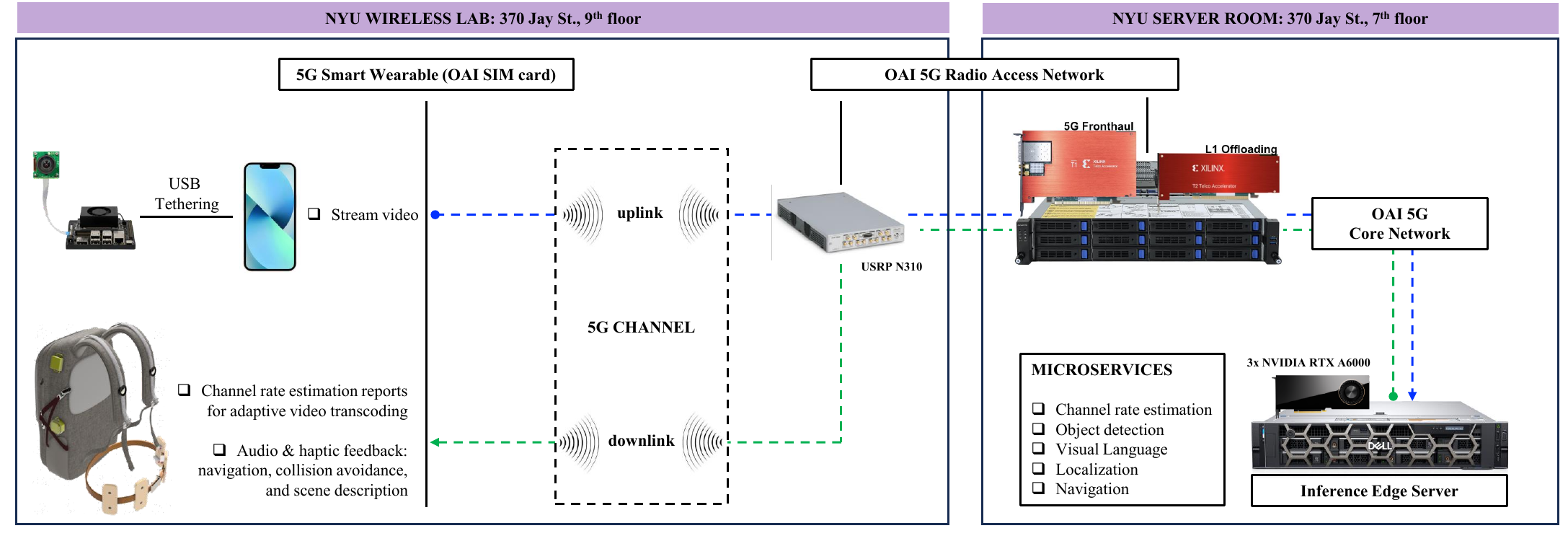}
  \caption{End-to-end 5G real-time adaptive video streaming platform.}
  \label{fig:testbed}
  \vspace{-4mm}
\end{figure*}

\subsection*{Adaptive video streaming platform}

The real-time video streaming platform consists of a \emph{client} and a \emph{server}. The client runs on an NVIDIA Jetson board, which connects to a 5G-enabled smartphone for cellular connectivity, as shown in Fig.~\ref{fig:testbed}. Primarily, the client initiates an uplink video stream to the server using \gls{rtp} and receives a sequence of outputs from the microservices introduced above. The reception of such information is important to unleash the full potential of \VISION\ and assist \gls{pblv} through haptic and audio feedback. Importantly, to ensure seamless performance of these assistive microservices, video streaming must take into account link rate fluctuations that are related but not limited to network congestion, poor wireless channel quality, and the number of users simultaneously accessing the same wireless resources. As such, our video streaming client implements an adaptive mechanism that enables video encoder adjustments based on channel quality. Specifically, the client periodically receives link rate estimates from the server and then uses the REBERA algorithm \cite{REBERA} to predict the link rate in the next time step based on previous observations and accordingly set the appropriate encoder rate. In addition, the client adapts the video streaming resolution according to the set video rate. As we found in \cite{azzino}, lowering the video resolution when the target video rate is below a certain threshold improves object detection performance. 

Navigation and \gls{vlm}-based scene description do not need to run at high frame rate due to their high computational load and relaxed latency requirements. A single high-quality video frame per second should suffice. Hence, we support multiple video-based services at different frame rates and spatial resolutions. In particular, when the predicted link rate is high, we send a single video stream with a high frame rate (30 fps) and high spatial resolution (1920$\times$1080). In this case, object detection will be performed for every frame, but navigation and \gls{vlm} will be performed only once every second, hence every 30 frames are received. When the predicted link quality degrades, we reduce the video resolution but keep the same frame rate so that object detection can still be performed frequently but with reduced accuracy. We further enable a secondary video stream with low frame rate (e.g. 1 fps), but high spatial resolution, to perform navigation and \gls{vlm} inference on high-quality and resolution images. As outlined in \cite{azzino}, high-resolution images enable the detection of objects far away (which would occupy a small region in an image and are more difficult to detect from low-resolution images). Sending high-resolution video when the network connectivity is strong allows the user to ``see'' objects far away and plan their trajectory properly. When the network link is weak, despite the reduced image resolution, we can still accurately detect nearby objects (which have a relatively larger image size and can still be easily detected in low-resolution images) that could pose a threat to \gls{pblv}. We choose to perform navigation and \gls{vlm} services only in high-resolution frames because our separate studies have found that increasing spatial resolution can significantly improve navigation accuracy \cite{yang2023distillation}. 
Effectively, the presence of a secondary video stream adds a second layer of adaptation based on the \gls{qos} requirement for a given microservice. 

As depicted in Fig.~\ref{fig:testbed}, the server runs on any computer system equipped with \glspl{gpu}. It receives the uplink video stream from the client and performs a series of services that are instrumental in supporting the full set of features expected for \VISION. Similarly as in~\cite{REBERA}, the server periodically computes an estimate of the link rate based on the incoming stream of video packets. Each estimate is sent back to the client for video adaptation as previously discussed. The server implements a pipeline to support real-time, high-frame rate object detection on the received video frames. In addition, the server hosts microservices for navigation and localization, and vision-language inference on high-resolution frames at a low frame rate. All these services run on parallel threads in order to distribute their computational requirements over different resources on the same machine. Furthermore, the server supports the reception of a secondary low-frame rate, high-resolution video stream when enabled by the client. As such, if this stream is present, microservices that have a lower frame rate requirement (i.e., navigation and localization, and \gls{vlm} instruction) enjoy high-quality video frames. If this stream is not running, the mentioned microservices use frames from the primary high-frame rate video stream.

Finally, client and server have been developed using GStreamer libraries \cite{gstreamer} to stream real-time video, and DeepStream \cite{nvidia-deepstream} for real-time object detection with YoLo running on NVIDIA's \glspl{gpu}. Socket and thread programming in Python has been used to enable parallel execution of microservices and link rate estimation.

\section{5G Experimental Platform}
\label{sec:testbed}

\begin{figure*}[t!]
 \newcommand\fheight{0.4\linewidth}
 \newcommand\fwidth{0.75\linewidth}
 \vspace{-3cm}
 \hspace*{\fill}%
 \hspace{0.5cm}\begin{subfigure}[b]{0.35\textwidth}
  \centering
  \includegraphics[width=0.98\linewidth]{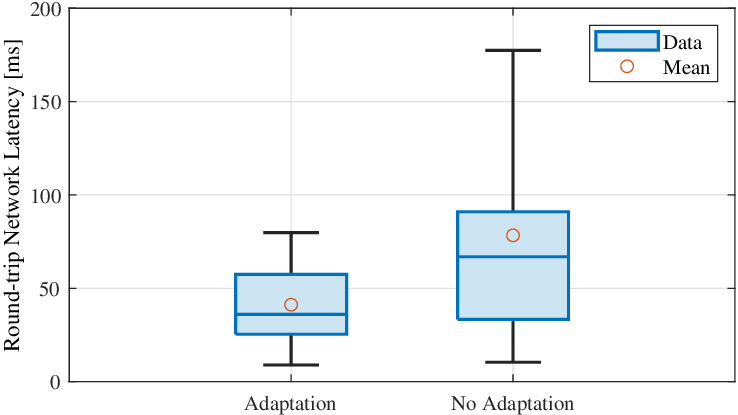}
  \caption{Box-plots for the round-trip 5G network latency.}
  \label{fig:box_network_latency}
  \vspace{-7.2cm}
 \end{subfigure}
 \hspace*{\fill}%
 \hspace{0.85cm}\begin{subfigure}[b]{0.35\textwidth}
  \centering
  \includegraphics[width=0.98\linewidth]{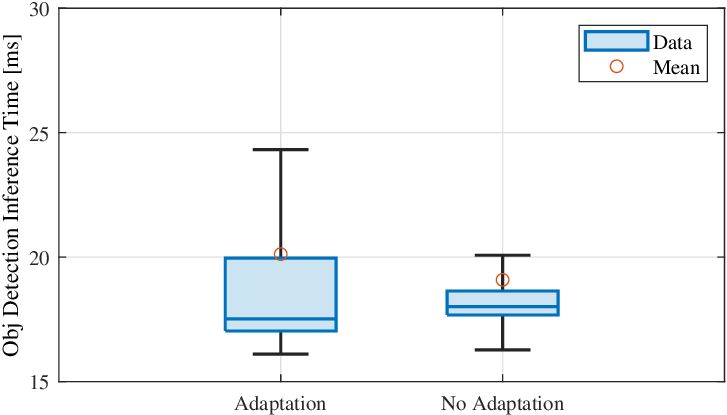}
  \caption{Box-plots for the object detection inference time.}
  \label{fig:box_inference_time}
  \vspace{-7.2cm}
 \end{subfigure}
 \hspace*{\fill}%
 \\
 \hspace*{\fill}%
 \hspace{0.6cm}\begin{subfigure}[b]{0.42\textwidth}
   \centering
   \input{figures/results_1_cdf_e2e_time_adap_vs_noadap}
   \caption{CDF of the end-to-end inference latency. It includes round-trip networking and object detection inference times.}
   \label{fig:cdf_e2e_time}
 \end{subfigure}
 \hspace*{\fill}
 \begin{subfigure}[b]{0.42\textwidth}
   \centering
   \input{figures/results_1_est_link_rate_sender_rate}
   \caption{Maximum link rate, estimated link rate, and encoder bitrate over the duration of the experiment.}
   \label{fig:link_rate_est_rate}
 \end{subfigure}
 \hspace*{\fill}%
 \caption{Results for the video streaming experiment on our platform. The blue box in the box plots represents the interquartile range between the 25th and 75th percentiles. The blue line represents the median.}
\vspace*{-5mm}
 \label{fig:all_results}
\end{figure*}

\subsection*{5G cellular stack}
A programmable 5G network was deployed using the widely adopted \gls{oai} stack~\cite{kaltenberger2020openairinterface}. \gls{oai} is an open-source project that implements \gls{3gpp} technology on general-purpose x86 computing hardware and \gls{cots} \gls{sdr} cards such as \glspl{usrp}. Importantly, the \gls{oai} framework includes both \gls{ran} and \gls{5gc}, enabling end-to-end 5G cellular functionalities. As shown in Fig.~\ref{fig:testbed}, a commercial phone gets programmable 5G connectivity using an \gls{oai} \gls{sim} card. The 5G \gls{ran} components are compiled on the host machine, whereas the \gls{5gc} is loaded on the same machine by launching several docker images that will jointly verify, authenticate, and subscribe each preregistered \gls{oai} device. Once connected to 5G, the user will be able to communicate with any accessible servers on the external data network. In our case, as discussed in further detail in the next section, we deployed a powerful inference server on the same dedicated network as \gls{5gc}, thus mimicking the performance of an edge server. Finally, a major benefit of using open-source cellular software is that key parameters can be logged and tracked at any network layer, allowing us to accurately profile the overall performance of our experiments.

\noindent 
\textbf{\gls{oai} code modification -- variable BWPs:} Critically, to assess the advantages of our adaptive video streaming technology, we have made modifications to the \gls{oai} \gls{gnb} scheduler code to trigger specific variations in channel capacity. To do so, we leverage a new feature in 5G, that is \gls{bwp}, which was introduced in \gls{3gpp} Release 15 to dynamically adapt the carrier bandwidth and the associated numerology for each user \cite{release15}. We programmed the scheduler in such a way that \gls{gnb} will modify the allocated \gls{bwp} configuration at specific time intervals.

\subsection*{Hardware}
\textbf{5G Network:} As shown in Fig.~\ref{fig:testbed}, the 5G network infrastructure comprises the following hardware: the \gls{gnb} \emph{Radio Unit} runs on the USRP N310, which connects to the Proxicast 4G/5G Aerial Antennas; the \gls{gnb} \emph{Baseband Unit} and the \gls{5gc} run on the \emph{Nautilus} machine, which is a \gls{hpc} server equipped with Xilinx T1 and T2 boards for L1 acceleration. 


\textbf{Client:} The client includes a Jetson Orin NX 16 GB that connects to a wide-angle camera. This device has 5G \gls{oai} connectivity through USB tethering via a Google Pixel 5A. 

\textbf{Server:} We use a powerful Dell Precision 7920 Rack Workstation with 3x NVIDIA RTX A6000 (48GB each) to host the video streaming server with \gls{ai} edge services for our platform. This machine shares the same rack as the Nautilus server in order to mimic a realistic edge computing environment.

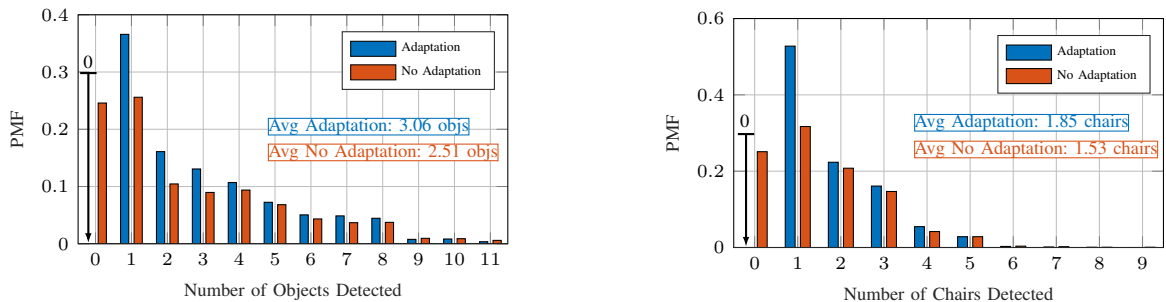
\begin{figure*}[t!]
 \newcommand\fheight{0.4\linewidth}
 \newcommand\fwidth{0.75\linewidth}
 \hspace*{\fill}%
 \begin{subfigure}[b]{0.42\textwidth}
   \centering
%
%
\definecolor{mycolor1}{rgb}{0.00000,0.44700,0.74100}%
\definecolor{mycolor2}{rgb}{0.85000,0.32500,0.09800}%
\begin{tikzpicture}

\pgfplotsset{every tick label/.append style={font=\scriptsize}}

\begin{axis}[%
width=\fwidth,
height=\fheight,
at={(0.758in,0.481in)},
scale only axis,
bar shift auto,
xmin=-0.485714285714286,
xmax=11.4857142857143,
xtick={ 0,  1,  2,  3,  4,  5,  6,  7,  8,  9, 10, 11},
xlabel style={font=\scriptsize\color{white!15!black}},
xlabel={Number of Objects Detected},
ymin=0,
ymax=0.4,
ylabel style={font=\scriptsize\color{white!15!black}},
ylabel={PMF},
axis background/.style={fill=white},
xmajorgrids,
ymajorgrids,
legend style={at={(0.97,0.8)}, font=\fontsize{5}{6}\selectfont, anchor=east, legend cell align=left, align=left, draw=white!15!black}
]
\addplot[ybar, bar width=0.229, fill=mycolor1, draw=black, area legend] table[row sep=crcr] {%
0	0\\
1	0.365795724465558\\
2	0.160926365795724\\
3	0.130641330166271\\
4	0.106888361045131\\
5	0.0724465558194774\\
6	0.0504750593824228\\
7	0.0486935866983373\\
8	0.0445368171021378\\
9	0.00771971496437055\\
10	0.00831353919239905\\
11	0.00356294536817102\\
};
\addplot[forget plot, color=white!15!black] table[row sep=crcr] {%
-0.485714285714286	0\\
11.4857142857143	0\\
};
\addlegendentry{Adaptation}

\addplot[ybar, bar width=0.229, fill=mycolor2, draw=black, area legend] table[row sep=crcr] {%
0	0.2458432304038\\
1	0.255938242280285\\
2	0.104513064133017\\
3	0.089667458432304\\
4	0.0938242280285036\\
5	0.0682897862232779\\
6	0.0433491686460808\\
7	0.0368171021377672\\
8	0.0374109263657957\\
9	0.00950118764845606\\
10	0.00890736342042755\\
11	0.00593824228028504\\
};
\addplot[forget plot, color=white!15!black] table[row sep=crcr] {%
-0.485714285714286	0\\
11.4857142857143	0\\
};
\addlegendentry{No Adaptation}

\node[fill=white, below right, align=left, inner sep=0, font=\scriptsize\color{mycolor1}, draw=mycolor1]
at (4.8,0.22) {Avg Adaptation: 3.06 objs};
\node[fill=white, below right, align=left, inner sep=0, font=\scriptsize\color{mycolor2}, draw=mycolor2]
at (4.8,0.175) {Avg No Adaptation: 2.51 objs};

\node[fill=white, below right, align=left, inner sep=0, font=\scriptsize\color{black}]
at (-0.35,0.33) {0};
\draw[<-|, thick, >={LaTeX[length=4pt,width=2.5pt]}] (-0.2,0) -- (-0.2,0.3);

\end{axis}
\end{tikzpicture}%
   \caption{Bars plot with the PMF of the total number of objects detected per frame.}
   \label{fig:cdf_objects_all}
 \end{subfigure}
 \hspace*{\fill}
 \begin{subfigure}[b]{0.42\textwidth}
   \centering
%
%
\definecolor{mycolor1}{rgb}{0.00000,0.44700,0.74100}%
\definecolor{mycolor2}{rgb}{0.85000,0.32500,0.09800}%
\begin{tikzpicture}

\pgfplotsset{every tick label/.append style={font=\scriptsize}}

\begin{axis}[%
width=\fwidth,
height=\fheight,
at={(0,0)},
scale only axis,
bar shift auto,
xmin=-0.485714285714286,
xmax=9.48571428571428,
xtick={0, 1, 2, 3, 4, 5, 6, 7, 8, 9},
xlabel style={font=\scriptsize\color{white!15!black}},
xlabel={Number of Chairs Detected},
ymin=0,
ymax=0.6,
ylabel style={font=\scriptsize\color{white!15!black}},
ylabel={PMF},
axis background/.style={fill=white},
xmajorgrids,
ymajorgrids,
legend style={at={(0.97,0.8)}, font=\fontsize{5}{6}\selectfont, anchor=east, legend cell align=left, align=left, draw=white!15!black}
]
\addplot[ybar, bar width=0.229, fill=mycolor1, draw=black, area legend] table[row sep=crcr] {%
0	0\\
1	0.527659574468085\\
2	0.223404255319149\\
3	0.160992907801418\\
4	0.0546099290780142\\
5	0.0283687943262411\\
6	0.00283687943262411\\
7	0.00141843971631206\\
8	0.000709219858156028\\
9	0\\
};
\addplot[forget plot, color=white!15!black] table[row sep=crcr] {%
-0.485714285714286	0\\
9.48571428571428	0\\
};
\addlegendentry{Adaptation}

\addplot[ybar, bar width=0.229, fill=mycolor2, draw=black, area legend] table[row sep=crcr] {%
0	0.251063829787234\\
1	0.317021276595745\\
2	0.207801418439716\\
3	0.146808510638298\\
4	0.0418439716312057\\
5	0.0283687943262411\\
6	0.00354609929078014\\
7	0.00212765957446809\\
8	0.000709219858156028\\
9	0.000709219858156028\\
};
\addplot[forget plot, color=white!15!black] table[row sep=crcr] {%
-0.485714285714286	0\\
9.48571428571428	0\\
};
\addlegendentry{No Adaptation}

\node[fill=white, below right, align=left, inner sep=0, font=\scriptsize\color{mycolor1}, draw=mycolor1]
at (3.7,0.35) {Avg Adaptation: 1.85 chairs};
\node[fill=white, below right, align=left, inner sep=0, font=\scriptsize\color{mycolor2}, draw=mycolor2]
at (3.7,0.28) {Avg No Adaptation: 1.53 chairs};

\node[fill=white, below right, align=left, inner sep=0, font=\scriptsize\color{black}]
at (-0.35,0.35) {0};
\draw[<-|, thick, >={LaTeX[length=4pt,width=2.5pt]}] (-0.2,0) -- (-0.2,0.3);

\end{axis}
\end{tikzpicture}%
   \caption{Bars plot with the PMF of the number of \emph{chairs} detected per frame.}
   \label{fig:cdf_chair}
 \end{subfigure}
 \hspace*{\fill}%
 \caption{Object detection results for the video streaming experiment on our platform.}
 \label{fig:object_detection}
 \vspace{-3mm}
\end{figure*}

\section{Results}
\label{sec:results}

In this section, we discuss the experimental results obtained using the 5G platform introduced above. Our goal is to test our adaptive video streaming platform along with the microservices enabled for \gls{pblv}, and to measure its overall performance. Critically, we compared the results of each microservice with and without video adaptation enabled. Without video adaptation, we fix the encoder bitrate to 20 Mbps, the frame rate to 30 fps, and the video resolution to 1920$\times$1080.

We recorded a high-bitrate and resolution video in which we walk from our wireless lab to the kitchen on the same floor. We placed a set of potentially hazardous obstacles such as office chairs along the route. Then we walked from the lab room to the kitchen, recording approximately 60 seconds of video. We then emulated the streaming of this video under time-varying network conditions and invoked the three \gls{ai} services on the decoded video on the inference edge server. A pre-recorded video allowed us to compare the performance of the multiple \gls{ai} services with video adaptation vs. without video adaptation under the same controlled network conditions and using the exact same video frames. We tethered the Jetson board to the Google Pixel to gain 5G \gls{oai} connectivity. The Google Pixel, which was wirelessly connected to \gls{gnb}, did not move due to coverage constraints in our 5G frontend and therefore experienced stable channel quality throughout the duration of the experiment. To mimic 5G link rate fluctuations, we modified \gls{bwp} every $\approx$ 10 seconds, as reported in Table~\ref{tab:params}, and as shown in Fig.~\ref{fig:link_rate_est_rate} by the dashed yellow line representing the maximum link rate.\footnotemark
\afterpage{\footnotetext{Note: the maximum link rate is lower than the capacity expected for each \gls{bwp} because the uplink performance is limited by the computational resources of our Nautilus server. To overcome this limitation, we are integrating FPGA-based LDPC acceleration enabled by the Xilinx T1 card.}}

\begin{table}[t!]
    \fontsize{8pt}{8pt}\selectfont
    \centering
    \caption{Main parameters for the 5G \gls{oai} experiments. Relatively to the TDD configuration, \textbf{D} indicates a downlink slot, \textbf{U} stands for uplink, and \textbf{S} is a special slot with a mixture of 6 downlink symbols and 4 uplink symbols.}
    \begin{tabular}{|>{\raggedright}m{1.6cm}|>{\raggedright}m{1.8cm}|>{\raggedright}m{2.3cm}|>{\raggedright}m{1.1cm}|}
    \hline
    \textbf{Parameter} & \textbf{Value} & \textbf{Parameter} & \textbf{Value}
    \tabularnewline \hline
    Frequency carrier & 2593.350 MHz & BWP 1 channel bw & 40 MHz
    \tabularnewline \hline
    TDD configuration & DSUUU & BWP 2 channel bw & 20 MHz
    \tabularnewline \hline    
    Max encoder bitrate & 20 Mbps & BWP 3 channel bw& 10 MHz
    \tabularnewline \hline    
    Video resolution & 1920x1080 & BWP 1, 2, 3 SCS & 30 KHz
    \tabularnewline \hline    
    \end{tabular}
    \label{tab:params}
    \vspace{-3mm}
\end{table}

A collection of experimental results is illustrated in Fig.~\ref{fig:all_results}. In particular, Fig.~\ref{fig:box_network_latency} shows box plots of the round-trip 5G network latency. Enabling adaptation, the average and median are approximately 40 ms, whereas the interquartile range is $\approx$ 30 ms. On the other hand, without adaptation, the network latency is consistently higher, with an average of 80 ms, and a median of $\approx$ 70 ms. Besides, the interquartile range and the maximum round-trip are higher. This indicates that without adaptation, latency becomes unpredictable, leading to higher jitter values and worse video quality. Fig.~\ref{fig:box_inference_time} suggests that the object detection inference time does not vary significantly in the two cases, as both average values are close to each other (i.e., roughly 20 ms). Without adaptation, the average inference time is slightly lower because fewer candidates are detected given the lower average video quality. Fig.~\ref{fig:cdf_e2e_time} shows the \gls{cdf} of the end-to-end inference latency computed as the sum of the round-trip network latency and the average object detection inference time. As we can see, with adaptation, we can clearly satisfy the 100 ms \gls{qos} requirement (as in \cite{azzino}) for object detection all the time. Instead, without adaptation, 40\% of the time, we cannot meet the requirement. The results show that our adaptive video streaming platform can provide better performance for object detection compared to an implementation without adaptation.

Fig.~\ref{fig:link_rate_est_rate} shows the predicted channel link rate and the adapted encoder bitrate during the video adaptation experiment. Note that the encoder rate without adaptation would be fixed to the maximum value of 20 Mbps. We could transmit the video at lower bitrate (i.e. underestimating the channel condition); nonetheless, this will result in consistent degraded performance for all the microservices due to poor image quality. From the figure we can see that the encoder bitrate closely follows the predicted link rate. In addition, the proximity between the prediction and the actual channel link rate depends on the amount of data sent on the uplink. As such, when the encoder bitrate is capped at its maximum value of 20 Mbps, the predicted link rate is consistently lower compared to the actual maximum channel link rate.

YoLoV7 \cite{yolov7} can track objects from 80 classes; however, in this experiment, we considered a subset of them consisting of eight typical objects that can be found in an office environment. From the \gls{pmf} of Fig.~\ref{fig:cdf_objects_all}, we can see that a higher number of objects per frame is detected with adaptation. Moreover, from Fig.~\ref{fig:cdf_chair}, more chairs can be detected with adaptation. Without adaptation, there is a non-negligible probability that no objects or chairs are detected. Therefore, video adaptation helps to increase our \VISION wearable object detection capabilities.

\begin{table*}
    \renewcommand{\arraystretch}{1.9}
    \centering
    \scriptsize
    \setlength{\arrayrulewidth}{0.85pt}
    \caption{Navigation and localization and \gls{vlm} inference results.}
    \begin{tabular}{|p{0.2\textwidth}|p{0.15\textwidth}|p{0.26\textwidth}|p{0.25\textwidth}|}
        \hline 
        \rowcolor{purple!15} \textbf{Computational flow at edge} & \textbf{Item} &
        \textbf{Adaptation} & \textbf{No Adaptation} \\
        \hline
        \multirow{6}{0.2\textwidth}{
        \begin{minipage}{0.2\textwidth}
            \vfill
            \includegraphics[width=0.9\linewidth, height=7.5cm]{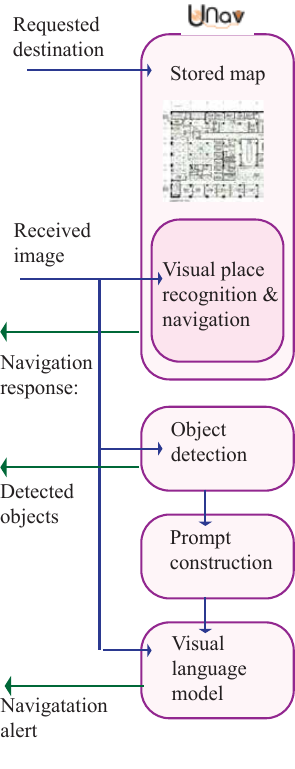} 
            \vfill
        \end{minipage}
            
        }
        & Requested destination & \multicolumn{2}{c|}{``Kitchen room of 9th floor at 370 Jay St."} 
        \\ \hhline{|~|-|-|-} 
       &  Received image at the edge along with detected objects
        by Yolo running at the edge &
        \begin{minipage}{0.24\textwidth}
            \centering
            \begin{subtable}{\linewidth}
                \vspace{0.2cm}
                \includegraphics[width=\linewidth]{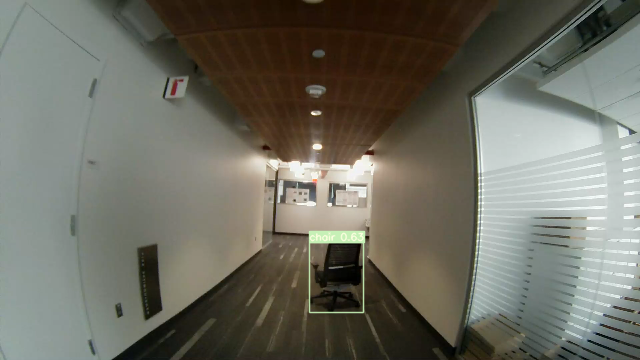}
                \vspace{0.15cm}
            \end{subtable}
        \end{minipage} &
        \begin{minipage}{0.24\textwidth}
            \centering
            \begin{subtable}{\linewidth}
                \vspace{0.2cm}
                \includegraphics[width=\linewidth]{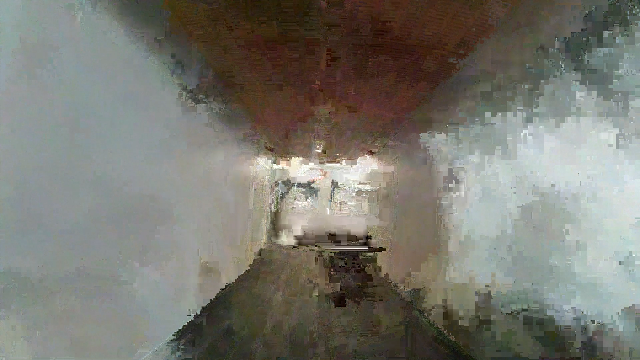}
                \vspace{0.15cm}
            \end{subtable}
        \end{minipage} \\ \hhline{|~|-|-|-}
       &  Detected objects & Chair & None \\ 
       \hhline{|~|-|-|-}
       &\cellcolor{yellow!30} Navigation response to user &\cellcolor{yellow!30} ``Go straight to 12 o'clock and walk 15 feet. Then turn right." &\cellcolor{yellow!30} ``No path to destination" \\
        \hhline{|~|-|-|-}
       & {\cellcolor{cyan!20}} Engineered prompt (automatically generated, not provided by the user) & 
        {\cellcolor{cyan!20}}"There appears to be a chair in the scene. Is the chair a threat to a user with blindness who wants to walk forward? If yes, in which direction should the user go to avoid it? (avoid mentioning blindness in your answer)" & 
        {\cellcolor{cyan!20}}"Is there any object that can be a threat to a user with blindness who wants to walk forward? If yes, in which direction should the user go to avoid it?" (avoid mentioning blindness in your answer) \\ \hhline{|~|-|-|-}
        & Alert{\cellcolor{cyan!20}} given to user &
        {\cellcolor{cyan!20}} "Yes, the chair is a threat to a user who wants to walk forward. To avoid the chair, the user should go to the left side of the hallway. The chair is located on the right side of the hallway." 
        & {\cellcolor{cyan!20}}"No, there is no object that can be a threat to a user who wants to walk forward in this image. The user can simply walk forward without any obstacles." \\
        \hline
    \end{tabular}    
    \vspace{-2mm}
    \label{tab:navvlm}
\end{table*}

For navigation and localization, we mapped the entire 9th floor at 370 Jay Street. For vision-language processing based on \gls{vlm}, we deployed InstructBLIP as in \cite{instructblip}. Table~\ref{tab:navvlm} shows the importance of video adaptation to allow for precise inference in each microservice. In the case of adaptation, the video frame depicted is received from the low-frame rate, high-resolution secondary video stream, which is activated when the predicted link rate falls to 5 Mbps. Highlighted in yellow in the table, we have the output of UNav with and without adaptation based on the corresponding video frame as input. As can be seen, with adaptation, the output corresponds exactly to the correct navigation information to reach the destination; without adaptation, due to the poor image quality, UNav cannot localize the user, thus not finding any path to the destination. 

Highlighted in cyan in Table~\ref{tab:navvlm}, we show the results of the \gls{vlm} inference. In the case of adaptation, YoLo recognizes the \emph{chair} in the frame; hence we can use object detection results to create a specific prompt to input into the model, as reported in the table. With adaptation, InstructBLIP can assess the risks associated with the detected object along the path of \gls{pblv}. However, without adaptation, we are forced to ask a more generic question to the model, since the chair is not detected, which ultimately cannot capture the risk associated with the obstacle along the route. This is just one possible usage of the \gls{vlm} assistance microservice (assessing environmental risks and complementing object detection). Other possible usages include, as an example, scene description.

\section{Conclusions}
\label{sec:conclusion}

The adaptive video streaming platform presented in this paper aims to increase the performance of an ensemble of microservices based on \gls{ai} offered to \VISION, a smart wearable that helps \gls{pblv} in their daily challenges. To achieve that, we propose to (1) leverage edge computing resources to increase performance and minimize latency along with battery consumption, and (2) utilize adaptive video streaming strategies that maximize the quality of the video frames fed into the \gls{ai}-based microservices. This framework has been validated by means of real-world experiments. To do so, we built a testbed using \gls{oai}, which is an open-source implementation of 5G. This platform allowed us to derive real-world performance metrics such as end-to-end latency and visual processing accuracy. Overall, our results show the benefits of our proposed architecture both in terms of accuracy and aggregate latency --network plus inference--, which are particularly critical in the context of assistive technology. 
 
\vspace{-2mm}
\bibliographystyle{IEEEtran}
\bibliography{bibl}

\begin{thebibliography}{10}
\providecommand{\url}[1]{#1}
\csname url@samestyle\endcsname
\providecommand{\newblock}{\relax}
\providecommand{\bibinfo}[2]{#2}
\providecommand{\BIBentrySTDinterwordspacing}{\spaceskip=0pt\relax}
\providecommand{\BIBentryALTinterwordstretchfactor}{4}
\providecommand{\BIBentryALTinterwordspacing}{\spaceskip=\fontdimen2\font plus
\BIBentryALTinterwordstretchfactor\fontdimen3\font minus
  \fontdimen4\font\relax}
\providecommand{\BIBforeignlanguage}[2]{{%
\expandafter\ifx\csname l@#1\endcsname\relax
\typeout{** WARNING: IEEEtran.bst: No hyphenation pattern has been}%
\typeout{** loaded for the language `#1'. Using the pattern for}%
\typeout{** the default language instead.}%
\else
\language=\csname l@#1\endcsname
\fi
#2}}
\providecommand{\BIBdecl}{\relax}
\BIBdecl

\bibitem{WHO282}
WHO, ``{V}isual impairment and blindness. {WHO} fact sheet, 282. {G}eneva.''
  2010.

\bibitem{bourne2013causes}
R.~R. Bourne, G.~A. Stevens, R.~A. White, J.~L. Smith, S.~R. Flaxman, H.~Price,
  J.~B. Jonas, J.~Keeffe, J.~Leasher, K.~Naidoo \emph{et~al.}, ``Causes of
  vision loss worldwide, 1990--2010: a systematic analysis,'' \emph{The lancet
  global health}, vol.~1, no.~6, pp. e339--e349, 2013.

\bibitem{sherrod2014association}
C.~E. Sherrod, S.~Vitale, K.~D. Frick, and P.~Y. Ramulu, ``Association of
  vision loss and work status in the united states,'' \emph{Jama
  Ophthalmology}, vol. 132, no.~10, pp. 1239--1242, 2014.

\bibitem{arc-1}
A.~Boldini, A.~L. Garcia, M.~Sorrentino, M.~Beheshti, O.~Ogedegbe, Y.~Fang,
  M.~Porfiri, and J.-R. Rizzo, ``An inconspicuous, integrated electronic travel
  aid for visual impairment,'' \emph{ASME Letters in Dynamic Systems and
  Control}, vol.~1, no.~4, p. 041004, 2021.

\bibitem{arc-8}
A.~Boldini, J.-R. Rizzo, and M.~Porfiri, ``A piezoelectric-based advanced
  wearable: obstacle avoidance for the visually impaired built into a
  backpack,'' in \emph{Nano-, Bio-, Info-Tech Sensors, and 3D Systems IV}, vol.
  11378.\hskip 1em plus 0.5em minus 0.4em\relax International Society for
  Optics and Photonics, 2020, p. 1137806.

\bibitem{azzino}
Z.~Yuan, T.~Azzino, Y.~Hao, Y.~Lyu, H.~Pei, A.~Boldini, M.~Mezzavilla,
  M.~Beheshti, M.~Porfiri, T.~E. Hudson, W.~Seiple, Y.~Fang, S.~Rangan,
  Y.~Wang, and J.-R. Rizzo, ``Network-aware 5g edge computing for object
  detection: Augmenting wearables to “see” more, farther and faster,''
  \emph{IEEE Access}, vol.~10, pp. 29\,612--29\,632, 2022.

\bibitem{intro_ta_1}
L.~Liu, H.~Li, and M.~Gruteser, ``Edge assisted real-time object detection for
  mobile augmented reality,'' in \emph{The 25th Annual International Conference
  on Mobile Computing and Networking}, ser. MobiCom '19, 2019.

\bibitem{REBERA}
E.~Kurdoglu, Y.~Liu, Y.~Wang, Y.~Shi, C.~Gu, and J.~Lyu, ``Real-time bandwidth
  prediction and rate adaptation for video calls over cellular networks,'' in
  \emph{Proceedings of the 7th International Conference on Multimedia Systems},
  ser. MMSys '16, 2016.

\bibitem{yolov7}
C.-Y. Wang, A.~Bochkovskiy, and H.-Y.~M. Liao, ``{YOLOv7}: Trainable
  bag-of-freebies sets new state-of-the-art for real-time object detectors,''
  \emph{arXiv preprint arXiv:2207.02696}, 2022.

\bibitem{yang2022unav}
A.~Yang, M.~Beheshti, T.~E. Hudson, R.~Vedanthan, W.~Riewpaiboon,
  P.~Mongkolwat, C.~Feng, and J.-R. Rizzo, ``{UNav: An
  Infrastructure-Independent Vision-Based Navigation System for People with
  Blindness and Low Vision},'' \emph{Sensors}, vol.~22, no.~22, p. 8894, 2022.

\bibitem{lowry2015visual}
S.~Lowry, N.~S{\"u}nderhauf, P.~Newman, J.~J. Leonard, D.~Cox, P.~Corke, and
  M.~J. Milford, ``{Visual place recognition: A survey},'' \emph{ieee
  transactions on robotics}, vol.~32, no.~1, pp. 1--19, 2015.

\bibitem{instructblip}
W.~Dai, J.~Li, D.~Li, A.~M.~H. Tiong, J.~Zhao, W.~Wang, B.~Li, P.~Fung, and
  S.~Hoi, ``Instructblip: Towards general-purpose vision-language models with
  instruction tuning,'' 2023.

\bibitem{vicuna2023}
\BIBentryALTinterwordspacing
W.-L. Chiang, Z.~Li, Z.~Lin, Y.~Sheng, Z.~Wu, H.~Zhang, L.~Zheng, S.~Zhuang,
  Y.~Zhuang, J.~E. Gonzalez, I.~Stoica, and E.~P. Xing, ``Vicuna: An
  open-source chatbot impressing gpt-4 with 90\%* chatgpt quality,'' March
  2023. [Online]. Available: \url{https://lmsys.org/blog/2023-03-30-vicuna/}
\BIBentrySTDinterwordspacing

\bibitem{yang2023distillation}
A.~Yang, Y.~Wang, J.-R. Rizzo, and C.~Feng, ``Distillation improves visual
  place recognition for low-quality queries,'' 2023.

\bibitem{gstreamer}
{GStreamer}, ``{GStreamer Library},'' \url{https://gstreamer.freedesktop.org/}.

\bibitem{nvidia-deepstream}
{NVIDIA}, ``{DeepStream SDK},''
  \url{https://developer.nvidia.com/deepstream-sdk}.

\bibitem{kaltenberger2020openairinterface}
F.~Kaltenberger, A.~P. Silva, A.~Gosain, L.~Wang, and T.-T. Nguyen,
  ``Openairinterface: Democratizing innovation in the 5g era,'' \emph{Computer
  Networks}, vol. 176, p. 107284, 2020.

\bibitem{release15}
{3GPP}, ``{Release 15},''
  \url{https://portal.3gpp.org/desktopmodules/Specifications/SpecificationDetails.aspx?specificationId=3389}.

\end{thebibliography}

\end{document}